\documentclass[final]{IEEEtran}
\usepackage{cite}
\usepackage[cmex10]{amsmath}
\usepackage{amsfonts}   
\usepackage{amssymb}    
\usepackage{verbatim}
\usepackage{latexsym}
\usepackage{graphicx}
\usepackage{epstopdf}

\def\argmin{\mathop{\operator@font argmin}}
\def\argmax{\mathop{\operator@font argmax}}
\makeatother

\newcommand{\bea}{\begin{array}}
\newcommand{\ena}{\end{array}}
\newcommand{\beq}{\begin{equation}}
\newcommand{\enq}{\end{equation}}

\newcommand{\beqa}{\begin{eqnarray}}
\newcommand{\enqa}{\end{eqnarray}}

\newcommand{\beqan}{\begin{eqnarray*}}
\newcommand{\enqan}{\end{eqnarray*}}

\newcommand{\AL}{\begin{enumerate}}
\newcommand{\ALE}{\end{enumerate}}

\def\addots{\mathinner{
    \mkern1mu\raise0pt\vbox{\kern7pt\hbox{.}}
    \mkern2mu\raise4pt\hbox{.}
    \mkern2mu\raise7pt\hbox{.}
    \mkern1mu}}

\def\sddots{\mathinner{
    \mkern.8mu\raise7pt\hbox{.}
    \mkern.8mu\raise4pt\hbox{.}
    \mkern.8mu\raise0pt\vbox{\kern7pt\hbox{.}}
    \mkern1mu}}

\def\saddots{\mathinner{
    \mkern.2mu\raise0pt\vbox{\kern7pt\hbox{.}}
    \mkern.2mu\raise4pt\hbox{.}
    \mkern.2mu\raise7pt\hbox{.}
    \mkern1mu}}

\makeatletter
\def\setboxz@h{\setbox\z@\hbox}
\def\wdz@{\wd\z@}
\def\boxz@{\box\z@}
\def\underset#1#2{\binrel@{#2}%
  \binrel@@{\mathop{\kern\z@#2}\limits_{#1}}}
\def\binrel@#1{\begingroup
  \setboxz@h{\thinmuskip0mu
    \medmuskip\m@ne mu\thickmuskip\@ne mu
    \setbox\tw@\hbox{$#1\m@th$}\kern-\wd\tw@
    ${}#1{}\m@th$}%
  \edef\@tempa{\endgroup\let\noexpand\binrel@@
    \ifdim\wdz@<\z@ \mathbin
    \else\ifdim\wdz@>\z@ \mathrel
    \else \relax\fi\fi}%
  \@tempa
}
\let\binrel@@\relax%
\makeatother

\def\circone{
    {\ooalign{
   \smash{\hskip3.0pt\raise0pt\hbox{\small $1$}}\vphantom{}\crcr
   \hbox{$\bigcirc$}
    }}
    }
\def\circtwo{
    {\ooalign{
   \smash{\hskip3.0pt\raise0pt\hbox{\small $2$}}\vphantom{}\crcr
   \hbox{$\bigcirc$}
    }}
    }
\def\circthree{
    {\ooalign{
   \smash{\hskip3.0pt\raise0pt\hbox{\small $3$}}\vphantom{}\crcr
   \hbox{$\bigcirc$}
    }}
    }

\def\sqplus{\mathbin{
    {\ooalign{\hfil\raise.3ex\hbox{\scriptsize
    +}\hfil\crcr\mathhexbox274\crcr\mathhexbox275}}
    }}
\def\sqminus{\mathbin{
    {\ooalign{\hfil\raise.3ex\hbox{\scriptsize
    --}\hfil\crcr\mathhexbox274\crcr\mathhexbox275}}
    }}

\def\IC{{
   \mathord{
      \hbox to 0em{
     \hskip-4pt
         \ooalign{
       \smash{\hskip1.9pt\raise2.6pt\hbox{$\scriptscriptstyle |$}}\crcr
       \smash{\hbox{\rm\sf C}} }
     \hidewidth}
      \phantom{\hbox{\rm\sf C}}
} }}
\def\IN{
    {\ooalign{
   \smash{\hskip2.2pt\raise1.5pt\hbox{$\scriptscriptstyle |$}}\vphantom{}\crcr
   \hbox{\sf N}
    }}
    }
\def\IZ{
    {\ooalign{
   \smash{\hskip1.9pt\raise0pt\hbox{$\sf Z$}}\vphantom{}\crcr
   \hbox{\sf Z}
    }}
    }
\def\IR{
    {\ooalign{
   \smash{\hskip2.2pt\raise1.5pt\hbox{$\scriptscriptstyle |$}}\vphantom{}\crcr
   \smash{\hskip2.2pt\raise3.3pt\hbox{$\scriptscriptstyle |$}}\vphantom{}\crcr
   \hbox{\sf R}
    }}
    }

\iffalse
    \DeclareMathAlphabet{\mathcmb}{OT1}{cmr}{bx}{n}

\else
    \usepackage{amsbsy}

\fi

\newcommand{\SI}{\begin{indlist}}
\newcommand{\EI}{\end{indlist}}

\newcommand{\DL}{\begin{dashlist}}
\newcommand{\DLE}{\end{dashlist}}

\usepackage{color}
\usepackage{multirow}
\usepackage{comment}
\usepackage{ragged2e}
\usepackage{booktabs}
\usepackage{diagbox}
\usepackage{bbding}
\usepackage{setspace}
\usepackage{subcaption}

\begin{document}
\title{Carrier Frequency Offset Estimation for OCDM with Null Subchirps}
\author{Sidong Guo,~\IEEEmembership{Student Member,~IEEE}, Yiyin Wang,~\IEEEmembership{Member,~IEEE,}
      and Xiaoli Ma,~\IEEEmembership{Fellow,~IEEE,}}


\maketitle

\begin{abstract}

In this paper, we investigate the carrier frequency offset (CFO) identifiability problem in orthogonal chirp division multiplexing (OCDM) systems. We propose a transmission scheme by inserting consecutive null subchirps. A CFO estimator is accordingly developed to achieve a full acquisition range. 
We further demonstrate that the proposed transmission scheme not only help to resolve CFO identifiability issues but also enable multipath diversity for OCDM systems. Simulation results corroborate our theoretical findings. 
\end{abstract}

\begin{IEEEkeywords}
CFO, OCDM, multipath diversity, identifiability
\end{IEEEkeywords}

\section{Introduction}
Orthogonal chirp division multiplexing (OCDM) is a multi-carrier modulation scheme that uses orthogonal chirp waveforms to carry information symbols. OCDM originates from paraxial optics and is first applied to wireless communications \cite{ouyang2016orthogonal} \cite{ouyang2017chirp}. Compared to the currently dominant modulation scheme, i.e., orthogonal frequency division multiplexing (OFDM), OCDM is more robust against frequency-selective channels, enjoys lower bit error rate (BER) under various equalization methods, and performs better against burst interferences. Gaining more attractions recently, OCDM also sees increased interests in the research of multi-user OCDM, channel and CFO estimation \cite{omar2019designing, ouyang2018robust, zhang2021channel, de2022joint, de2022discrete}.

In an OCDM system, the inverse discrete Fresnel transform (IDFnT) is used as the transform kernel to spread each information symbol over the entire bandwidth and the whole symbol block. Thus, OCDM has the double-spreading feature. A transform basis being non-sinusoidal may enjoy certain advantages against CFO. For instance, simulation results in \cite{omar2021performance} indicate that OCDM with a minimum mean square error (MMSE) equalizer or decision feedback equalizer (DFE) has a much better performance in terms of BER than OFDM over frequency-selective channels with CFO at medium signal-to-noise ratios (SNRs). It is also observed in \cite{trivedi2020low} that chirp-based OFDM systems outperforms traditional OFDM with the presence of CFO. 

The double-spreading property of chirp basis suggests that frequency-selective channels may have different impacts on receiver's ability to correctly identify CFO in OCDM compared to OFDM. The identifiability issue is well documented in the case of OFDM in \cite{ma2001non}. Owing to the sinusoidal kernel, OFDM converts a frequency-selective channel into a set of frequency-flat channels. However, channel nulls in the frequency domain can create multiple indistinguishable candidate CFOs in OFDM systems. Thus, it was proposed that judicially placed null subcarriers can restore CFO identifiability in OFDM, where channel nulls are sufficiently covered by null subcarriers \cite{ma2001non, ghogho2001optimized}. The optimality and identifiability of data-aided CFO estimation approaches for OFDM systems with training sequences are explored in \cite{gao2006identifiability, gao2008scattered}.


On the other hand, some preliminary efforts are directed towards handling CFO in OCDM. In \cite{zhang2021channel}, an excessive cyclic prefix (CP) is used to cross correlate with its counterpart in an OCDM block to estimate the CFO. In \cite{de2022joint}, a preamble-type synchronization and channel estimation method is applied to an OCDM system. Although both of them \cite{zhang2021channel,de2022joint} show decent accuracy, they require redundant CPs or symbols. They also face identifiability issues at a high normalized CFO. For instance, the CP-based approach \cite{zhang2021channel} has ambiguity in the CFO estimation, when the CFO exceeds half of the subchirp frequency spacing.

This paper resolves the CFO identifiability issue in OCDM 
with the help of the proposed transmission scheme, which inserts consecutive null subchirps. A CFO estimator is designed accordingly to achieve a full acquisition range of CFO. 
We further prove that the proposed transmission scheme not only help to address CFO identifiability issue, but also enable multipath diversity for OCDM systems. Simulation results verify the uniqueness of the proposed CFO estimator, and indicate that the BER performances of both the linear equalizers (LEs) and maximum likelihood estimator (MLE) benefit from the enabled multipath diversity.


Notations: Upper (lower) case bold letters are used to represent matrices (vectors). The notation \([\mathbf{A}]_{m,n}\) indicates the entry at the \(m\)th row and \(n\)th column of the matrix \(\mathbf{A}\), whereas \([\mathbf{A}]_{m,:}\) and \([\mathbf{A}]_{:,n}\) denotes the $m$th row and $n$th column of the matrix \(\mathbf{A}\), respectively. The identity matrix of size \(P\times P\) is denoted by \(\mathbf{I}_P\). The zero matrix of size \(M\times N\) is denoted by \(\mathbf{0}_{M \times N}\). The notations \((\cdot)^{T}\), \((\cdot)^\mathcal{H}\), \((\cdot)^{-1}\), \((\cdot)^\dagger\), and \((\cdot)^\ast\) represent transpose, conjugate transpose, inverse, Moore-Penrose inverse, and element-wise conjugate, respectively. We use \(E[\cdot]\) to denote the expectation with respect to (w.r.t.) all random variables within the brackets. 
\section{System Model}
In this section, we derive a baseband equivalent OCDM system model under frequency-selective channels with CFO, where the time-invariant frequency-selective channels are assumed with channel impulse responses (CIRs) \(h(l)\) of order \(L\), where \(l = 0,1,\dots,L\).

Consider the \(i\)th symbol block is composed of the $i$th data symbol block $\mathbf{s}(i)$ and null symbols. The null symbols in the Fresnel domain correspond to null subchirps in the time domain. The null symbols are added in a consecutive way. Thus, the $i$th symbol block of length $N$ is assembled as $\mathbf{T}_{\rm zp}\mathbf{s}(i)$, where $\mathbf{s}(i)$ is of length $K$ and \(\mathbf{T}_{\rm zp}=[\mathbf{I}_K\quad \mathbf{0}_{K\times (N-K)}]^{T}\).  We further assume that the covariance matrix of $\mathbf{s}(i)$ is given by $\mathbf{R}_{ss}=E[\mathbf{s}(i)\mathbf{s}^\mathcal{H}(i)] = E_s \mathbf{I}_K$, where $E_s$ is the symbol energy. The $i$th symbol block is modulated with the IDFnT. The resulting time-domain symbol block can be written as \(\mathbf{x}(i)=\mathbf{\Phi}^\mathcal{H}\mathbf{T}_{\rm zp}\mathbf{s}(i)\), where the DFnT matrix $\mathbf{\Phi}$ of size \(N \times N\) is defined as:
\begin{align}
    [\mathbf{\Phi}]_{m,n}=\frac{1}{\sqrt{N}}e^{-j\frac{\pi}{4}}\times & 
    \begin{cases}
        e^{j\frac{\pi}{N}(m-n)^2}, & N  = 0\,\,{\rm mod}\,\, 2, \\
        e^{j\frac{\pi}{N}(m+\frac{1}{2}-n)^2}, &N  = 1\,\, {\rm mod}\,\, 2,
    \end{cases} \nonumber \\
  & m,n = 1,\dots,N.\label{Phi}
\end{align}
The matrix $\mathbf{\Phi}$ is circulant and unitary, i.e., $\mathbf{\Phi}^\mathcal{H}\mathbf{\Phi}=\mathbf{\Phi}\mathbf{\Phi}^\mathcal{H}= \mathbf{I}_N$. Moreover, a cyclic prefix (CP) of length \(L\) is inserted in the time domain before $\mathbf{x}(i)$ to combat delay effects. 

The transmitted OCDM signal goes through a time-invariant frequency-selective channel. At the receiver side, we consider a normalized CFO \(w_o = 2\pi f_oT_s\) in the full range \([-\pi,\pi)\), where \(f_o\) is the CFO in Hz, and $T_s$ is the sampling period. After synchronization and removal of the CP, the received signal block $\mathbf{y}(i)$ of length $N$ is given by 
\begin{equation} \label{4}
    \mathbf{y}(i)=e^{j w_o(i(N+L)-N)}\mathbf{D}_N(w_o)\mathbf{H} \mathbf{\Phi}^\mathcal{H}\mathbf{T}_{\rm zp}\mathbf{s}(i)+\mathbf{n}(i),
\end{equation}
where \(\mathbf{D}_N(w_o)\) of size \(N \times N\) is the diagonal CFO matrix with the \(n\)th element in the diagonal given by \(e^{jw_o(n-1)}\), the first column of the circulant channel matrix \(\mathbf{H}\) is given by $\mathbf{h} = [h(0),\dots,h(L), \mathbf{0}_{N-L-1}^{T}]^{T}$, and \(\mathbf{n}(i)\) is the independent and identically distributed (i.i.d.) zero-mean additive white Gaussian noise (AWGN) vector with the covariance matrix $\sigma^2 \mathbf{I}_N$. 


Once the CFO estimate $\hat{w}_o$ is obtained, we compensate the CFO effect on the received signal block and achieve $\mathbf{r}(i) = e^{-j \hat{w}_o(i(N+L)-N)}\mathbf{D}_N^\mathcal{H}(\hat{w}_o)\mathbf{y}(i)$. Consequently, LEs \big(denoted by $\bf G_{(\cdot)}$\big) or non-linear equalizers can be applied to $\mathbf{r}(i)$. For example, the data symbols can be recovered using LEs as $\hat{\mathbf{s}}(i) = \mathbf{G}_{(\cdot)}\mathbf{r}(i)$, where the zero-forcing (ZF) or MMSE equalizers are given by
\begin{align}
\mathbf{G}_{\rm ZF}&=  \mathbf{B}^\dagger ,\label{ZF-FDE}\\
\mathbf{G}_{\rm MMSE}&=  \mathbf{B}^\mathcal{H} \left( \frac{\sigma^2}{E_{s}}\mathbf{I}_N+\mathbf{B}\mathbf{B}^\mathcal{H} \right)^{-1},\label{MMSE-FDE}
\end{align}
with \(\mathbf{B}={\mathbf{H}}\mathbf{\Phi}^\mathcal{H}\mathbf{T}_{\rm zp}\). 

\section{CFO Estimator and its Identifiability}

In this section, we develop an estimator for the true CFO \(w_0\) using the left null space (LNS) of the covariance matrix \(\mathbf{R}_{yy}=E[\mathbf{y}(i)\mathbf{y}^{\mathcal{H}}(i)]\). We first analyze the LNS of ${\mathbf{R}}_{yy}$. 
Moreover, a CFO estimator based on the LNS is proposed, and its uniqueness is proved.

\subsection{The Proposed CFO Estimator}

Based on \ref{4}, we derive the covariance matrix $\mathbf{R}_{yy}$ as follows, 
\begin{equation}
\mathbf{R}_{yy}
=\mathbf{D}_N(w_0)\mathbf{\Phi}^\mathcal{H}\mathbf{H} \mathbf{T}_{\rm zp}  \mathbf{R}_{ss} \mathbf{T}_{\rm zp}^\mathcal{H} \mathbf{H}^\mathcal{H}\mathbf{\Phi}\mathbf{D}_N^\mathcal{H}(w_0) \!+\!\sigma^2 \mathbf{I}_N, \label{Covariance1} 
\end{equation} where 
(\ref{Covariance1}) is achieved using the property of circulant matrices \( \mathbf{\mathbf{H}\Phi}^\mathcal{H}=\mathbf{\Phi}^\mathcal{H}\mathbf{H}\). 
Furthermore, the noiseless part of the covariance matrix is defined as
\begin{equation}\label{covariance_b}
\bar{\mathbf{R}}_{yy} =\mathbf{\Phi}^\mathcal{H}\mathbf{H} \mathbf{T}_{\rm zp}  \mathbf{R}_{ss} \mathbf{T}_{\rm zp}^\mathcal{H} \mathbf{H}^\mathcal{H}\mathbf{\Phi},
\end{equation} where $\bar{\mathbf{R}}_{yy}$ is of size $N \times N$ and ${\rm rank}(\mathbf{T}_{\rm zp}) = K$. Thus, the rank of $\bar{\mathbf{R}}_{yy}$ is upper-bounded by \( {\rm rank}(\bar{\mathbf{R}}_{yy}) \le {\rm min}\left({\rm rank}({\mathbf{H}}),{\rm rank}(\mathbf{T}_{\rm zp}) \right) < N\), according to Sylvester’s inequality. Hence, there exists the LNS of $\bar{\mathbf{R}}_{yy}$. 


The zero-padding (ZP) matrix $\mathbf{T}_{\rm zp}$ makes $\mathbf{H} \mathbf{T}_{\rm zp}$ a Toeplitz matrix of size $N \times K$. Recall that the multipath channel order is $L$. When $N-K > L$, the number of inserted null subchirps is greater than \(L\), and the last $N-K-L$ rows of $\mathbf{H} \mathbf{T}_{\rm zp}$ only contain zeros. As a result, we arrive at
\begin{equation}\label{LNS}
\mathbf{\Phi}^{\mathcal{H}}\mathbf{H} \mathbf{T}_{\rm zp} =\left[\mathbf{\Phi}^\ast_{\mathcal{S}}\,\, \mathbf{\Phi}^\ast_{\mathcal{N}} \right]
 \begin{bmatrix}
     \bar{\mathbf{H}} \\
     \mathbf{0}_{(N-K-L)\times K}
\end{bmatrix} ,
\end{equation} where $\mathbf{\Phi}^\ast_{\mathcal{S}} = \left[\mathbf{\Phi}^{\mathcal{H}}\right]_{:,1:K+L}$ of size $N \times (K+L)$, $\mathbf{\Phi}^\ast_{\mathcal{N}} = \left[\mathbf{\Phi}^{\mathcal{H}}\right]_{:,K+L+1:N}$ of size $N \times (N-K-L)$, and $\bar{\mathbf{H}} = \left[\mathbf{H}\right]_{1:K+L,1:K}$ of size $(K+L) \times K$. According to \ref{LNS}, we further prove that the null subchirps $\mathbf{\Phi}_{\mathcal{N}} = \left[\mathbf{\Phi}^{T}\right]_{:,K+L+1:N}$ is in the LNS of $\bar{\mathbf{R}}_{yy}$ as

\begin{align}
& \mathbf{\Phi}^T_{\mathcal{N}}\left[\mathbf{\Phi}^\ast_{\mathcal{S}}\,\, \mathbf{\Phi}^\ast_{\mathcal{N}} \right] 
\begin{bmatrix}
     \bar{\mathbf{H}} \\
     \mathbf{0}_{(N-K-L)\times K}
\end{bmatrix}  \nonumber\\
 = &\left[ \mathbf{0}_{(N-K-L)\times (K+L)}\,\, \mathbf{I}_{N-K-L} \right] \begin{bmatrix}
     \bar{\mathbf{H}} \\
     \mathbf{0}_{(N-K-L)\times K}
\end{bmatrix}  \nonumber\\
 = & \mathbf{0}_{(N-K-L) \times K}.
\end{align}
Define a set of indices $\mathcal{N}_{\phi} = \{K+L+1, \dots, N\}$, with this notation, the CFO $w_0$ can be estimated by using the LNS of $\bar{\mathbf{R}}_{yy}$ as
\begin{equation}\label{w_est}
\hat{w}_0= \mathop{\arg\min}\limits_{w}  J(w),
\end{equation} where the cost function $J(w)$ is defined as
\begin{equation}\label{Jw}
    J(w) =\sum_{k\in \mathcal{N}_{\phi}}\boldsymbol{\phi}_k^T\mathbf{D}_N^{-1}(w){\mathbf{R}}_{yy}\mathbf{ D}_N(w)\boldsymbol{\phi}_k^\ast, 
\end{equation} with $\boldsymbol{\phi}_k=\left[\mathbf{\Phi}^T\right]_{:,k}$ and $\boldsymbol{\phi}^\ast_k = \left[\mathbf{\Phi}^{\mathcal{H}}\right]_{:,k}$. Leveraging $\bar{\mathbf{R}}_{yy}$, the cost function \ref{Jw} can be rewritten as
\begin{align}
    J(w) 
=&\sum_{k\in \mathcal{N}_{\phi}}\boldsymbol{\phi}_k^T\mathbf{D}_N(w_0-w)\bar{\mathbf{R}}_{yy}\mathbf{ D}_N(w-w_0)\boldsymbol{\phi}_k^\ast \nonumber\\ 
&+ \sigma^2(N-K-L),
\end{align}
where the equivalent noise term $\sigma^2(N-K-L)$ is the result of $\sigma^2\sum_{k\in \mathcal{N}_{\phi}} \boldsymbol{\phi}_k^T \mathbf{D}_N^{-1}(w)\mathbf{I}_N\mathbf{D}_N(w)\boldsymbol{\phi}_k^\ast$, independent of the candidate CFO \(w\).
When \(w=w_0\), the result of \(\mathbf{D}_N(w_0-w)\) is an identity matrix, and the cost function is at a minimum, i.e., $J(w_0) = \sigma^2(N-K-L)$.


\subsection{CFO Identifiability}
The CFO identifiability issue arises for the CFO estimator \ref{w_est}, if the following necessary condition is fulfilled for some $w \neq w_0$
\begin{equation} \label{condition}
\left\{\boldsymbol{\phi}_{\widetilde{k}}^T\mathbf{D}_N(w_0-w) \right\}_{\widetilde{k} \in  \mathcal{N}_{\phi}} \subset \{\boldsymbol{\phi}_{k}^T\}_{k \in  \mathcal{N}_{\phi}}.
\end{equation} The condition in (\ref{condition}) for some $w \neq w_0$ results in multiple minimums. Based on the definition of $\mathbf{\Phi}$ in \ref{Phi}, it is clear that $\boldsymbol{\phi}_{k}$ is a chirp sequence and $\boldsymbol{\phi}_{k+\Delta k}$ is a circularly shifted version of $\boldsymbol{\phi}_{k}$, where $\boldsymbol{\phi}_{k+\Delta k}$ is obtained by circularly shifted $\boldsymbol{\phi}_{k}$ down (up) by $\Delta k$ samples, with $\Delta k$ being a non-negative (non-positive) integer. Hence, supposing that $k=K+L+\Delta k$, the LNS of the results of $\bar{\mathbf{R}}_{yy}$ can be rewritten as

\begin{equation}
\{\boldsymbol{\phi}_{k}^T\}_{k \in  \mathcal{N}_{\phi}} = \{\boldsymbol{\phi}_{K+L+\Delta k}^T\}_{\Delta k =1}^{N-K-L}.
\end{equation} Assume a block size of \(N\) with even value as an illustrative example. The $m$th entry of \(\boldsymbol{\phi}_{K+L+\Delta k}\) is given by:
\begin{align}\label{phi_1}
     &\left[\boldsymbol{\phi}_{K+L+\Delta k}\right]_m \\
     =& \frac{e^{-\frac{j\pi}{4}}}{\sqrt{N}}e^{j\frac{\pi}{N}(K+L+\Delta k-m)^2},\nonumber\\
     =& \frac{e^{-\frac{j\pi}{4}}}{\sqrt{N}}e^{j\frac{\pi}{N}(K+L-m)^2 +j\frac{\pi}{N}(\Delta k^2 + 2\Delta k(K+L)-2m\Delta k)},\nonumber\\
     &m=1,\dots, N, \Delta k = 1, \dots, N-K-L.\nonumber
\end{align}

On the other hand, the sequence $\boldsymbol{\phi}_{\widetilde{k}}^T\mathbf{D}_N(w_0-w)$ is also a shifted $\boldsymbol{\phi}_{\widetilde{k}}^T$. Let us define $\widetilde{k}=K+L+\Delta \widetilde{k}$, where $\Delta \widetilde{k} = 1, \dots, N-K-L$. 
Therefore, the $m$th entry of $\boldsymbol{\phi}_{\widetilde{k}}^T\mathbf{D}_N(w_0-w)=\boldsymbol{\phi}_{K+L+\Delta \widetilde{k}}^T\mathbf{D}_N(w_0-w)$ 
is given by
\begin{align}\label{phi_2}
 &\left[\boldsymbol{\phi}_{K+L+\Delta \widetilde{k}}^T\mathbf{D}_N(w_0-w)\right]_m \\
 =& \frac{e^{-\frac{j\pi}{4}}}{\sqrt{N}}e^{j(\frac{\pi}{N}(K+L+\Delta \widetilde{k}-m)^2+(w_0-w)m)},\nonumber\\
 =& \frac{e^{-\frac{j\pi}{4}}}{\sqrt{N}}e^{j\frac{\pi}{N}(K+L-m)^2}\nonumber\\
 &\times e^{j\frac{\pi}{N}(\Delta \widetilde{k}^2 + 2\Delta \widetilde{k}(K+L)-2m \Delta \widetilde{k}+\frac{N}{\pi}(w_0-w)m)},\nonumber\\
     &m=1,\dots, N,\Delta \widetilde{k} = 1, \dots, N-K-L.
     \nonumber
\end{align} Comparing \ref{phi_1} and \ref{phi_2}, we observe that $\boldsymbol{\phi}_{K+L+\Delta \widetilde{k}}^T\mathbf{D}_N(w_0-w)$ would belong to the LNS of $\bar{\mathbf{R}}_{yy}$ if and only if the two following conditions are fulfilled at the same time

\begin{align}
&\frac{\pi}{N}(\Delta \widetilde{k}^2 + 2\Delta \widetilde{k}(K+L))  \,\, {\rm mod} \,\, 2\pi \nonumber\\
=& \frac{\pi}{N}(\Delta k^2 + 2\Delta k(K+L)) \,\, {\rm mod} \,\, 2\pi,
\end{align} and 

\begin{align}
&\frac{2\pi}{N} m \Delta k \,\,  {\rm mod} \,\, 2\pi \nonumber\\
=& (\frac{2\pi}{N} m \Delta \widetilde{k} + (w_0-w)m) \,\, {\rm mod} \,\, 2 \pi
\end{align} 
for $\forall \Delta k$ and $\forall \Delta \widetilde{k} \in \{1,\dots,N-K-L\}$. Recall that $(w_0 - w) \in (-2\pi, 2\pi)$. It is easy to verify that the aforementioned two conditions can be satisfied simultaneously if and only if $\Delta k = \Delta \widetilde{k}$, and
$(w_0-w) \,\, {\rm mod}\,\, 2\pi =0$, i.e., $w_0= w$. Therefore, the CFO estimator employing the cost function \ref{Jw} has a unique minimum and the CFO identifiability issue is resolved. 

We conclude with the following proposition. 

\noindent \textbf{Proposition 1:} For the cost function \(J(w)\) in \ref{Jw} to have a unique minimum, the insertion of at least \(L+1\) consecutive null subchirps guarantees the CFO acqusition range $[-\pi, \pi)$ for an OCDM system under a multipath channel of order up to \(L\).  

It is worth noting that the spectral efficiency for the proposed method is \(\frac{N-L-1}{N+L}\), due to \(L+1\) unused sub-channels. 

\subsection{Error Performance}

Next, we explore the performance of the proposed OCDM transmission scheme with consecutive null subchirps (named OCDM-NSC) w.r.t. multipath diversity. A well known fact is that null subcarriers do not affect BER performance in OFDM w.r.t. multipath diversity. This is to say that the multipath diversity is always $1$ for OFDM systems due to the inherent diagonal structure of the equivalent channel. On the other hand, null subchirps affect the OCDM performance w.r.t. multipath diversity differently. The following proposition is herein. 

\noindent \textbf{Proposition 2:}
Suppose that the number of consecutive null subchirps is greater than or equal to the channel order $L$, i.e., $N-K \ge L$, the OCDM-NSC scheme achieves full multipath diversity by LEs and MLE.

\noindent \textbf{Proof:} 
There is a link between an OCDM with null subchirps and a ZP-only single carrier (SC) transmission scheme. It turns out that the demodulatd OCDM block with null subchirps also assumes the same mathematical form as the ZP-only SC block under multipath channels:
\begin{align}
    \mathbf{z}(i)  =&  \mathbf{\Phi}\mathbf{y}(i) \nonumber\\
=&\mathbf{\Phi}\mathbf{H}\mathbf{\Phi}^\mathcal{H}\mathbf{T}_{\rm zp} \mathbf{s}(i)  +\mathbf{n}(i) \nonumber\\
     =& \mathbf{H}\mathbf{T}_{\rm zp}\mathbf{s}(i)+\mathbf{n}(i), \nonumber
\end{align}
where \(\mathbf{H} = \mathbf{\Phi}\mathbf{H}\mathbf{\Phi}^\mathcal{H}\) by the property of the Fresnel matrix and the circulant matrix. It is observed in \cite{Wang_Optimality_2002} that ZP-only SC transmission with an equivalent tall toeplitz channel matrix \(\mathbf{H}\mathbf{T}_{zp}\) enables full multipath diversity with both LEs and MLE by providing a better channel matrix condition. OCDM-NSC after demodulation has the equivalent channel model. Therefore, this conclusion applies to OCDM-NSC.  \hfill $\blacksquare$
\section{Simulations}
In this section, we illustrate the performance results of the proposed method through simulations. In particular, the covariance matrix \(\mathbf{R}_{yy}\) is calculated empirically across \(N_b\) blocks as:
\begin{equation}
    \hat{\mathbf{R}}_{yy}=\frac{1}{N_b}\sum_{i=1}^{N_b}\mathbf{y}(i)\mathbf{y}^\mathcal{H}(i).
\end{equation} 
For all the simulations, we set that $N_b = 1000$, which is a sufficiently large empirical average to approximate \(\mathbf{R}_{yy}\). The SNR is defined as \(E_s/\sigma^2\).\\
To evaluate the performance of the CFO estimation and verify its identifiability, we define the mean square error (MSE) as \(\sum_{m=1}^{M}(w_{0}^{(m)}-\hat{w}_{0}^{(m)})^2/M\), where \(M\) is the number of Monte Carlo (MC) runs, and $w_{0}^{(m)}$ and $\hat{w}_{0}^{(m)}$ are the true and estimated CFO for the $m$th MC run, respectively. Multipath channels of order \(L=2\) with Rayleigh fading coefficients are employed. The block size is \(N=16\). We set \(K=12\) such that $N-K=4 > L +1$ null subchirps for the proposed CFO estimator. 
As our proposed CFO estimator is pilot-free, we opt to use the CP-based CFO estimator \cite{zhang2021channel} as baseline comparison instead of the pilot-aided Schmidl\&Cox method proposed in \cite{de2022joint}. The performance of the CP-based CFO estimator depends on 
the length of the excessive CP and the number of transmission blocks. Here we use an excessive CP of length \(L_{cp} = 4 > (L+1)\) and 
\(1000\) transmission blocks for the CP-based CFO estimator for a fair comparison.  
Although the CP-based CFO estimator \cite{zhang2021channel} is in a closed form, it faces the identifiability issue when $|w_0| > \pi/N$. The comparison results are shown in Fig.~1, where three cases of $w_0 \in [-0.05\pi,0.05\pi) $, $w_0 \in [-0.1\pi,0.1\pi)$, and $w_0 \in [-\pi, \pi)$ 
are presented, respectively. 
CP-based CFO estimator outperforms the proposed one, when $- \pi/N \le w_0 < \pi/N$ in the first case, due to the data-aided nature of cross-correlation based methods. In contrast, when $|w_0| > \pi/N$, the CFO identifiability issue arises for the CP-based CFO estimator, and the ambiguity error dominates its performance and results in an error floor. Since the identifiability issue is well addressed for the proposed CFO estimator, its MSE does not depend on the true CFO, and the estimator remains robust for the full range of the CFO, i.e., $w_0 \in [-\pi, \pi)$. 
We remark here that calculating the empirical covariance matrix in (18) requires a computational complexity of \(\mathcal{O}(N_bN^2)\). The proposed method incurs an additional cost in solving the minimization problem in (9), with computational complexity of \(\mathcal{O}(K N^2)\). Specifically, assuming a search candidate size of \(N_c\), complexity scales with \(\mathcal{O}(N_c K N^2)\). Thus, the total complexity of the proposed CFO estimator is \(\mathcal{O}((N_cK^2 + N_b)N^2)\). In comparison, the method in [5] has a complexity of \(\mathcal{O}(N_b(L_{cp}-L))\) for cross-correlation.
It is also worth pointing out that our proposed CFO estimator can work well in complement with the CP-based CFO estimator. The proposed method addresses the identifiability issue and achieve a coarse CFO compensation, and subsequently the CP-based method \cite{zhang2021channel} carries out a finer tuning. This resulting two-step CFO compensation scheme reaps both high accuracy and robustness against a full range CFO. 
\indent Next, we simulate the BER performance of the OCDM-NSC system with the proposed CFO estimator over multipath channels with Rayleigh fading coefficients. The QPSK modulation is adopted. The LEs specified in \eqref{ZF-FDE} and \eqref{MMSE-FDE} are applied, respectively. The true CFO $w_0$ is randomly generated following a uniform distribution in the range of $[-\pi, \pi)$ for each MC run. The simulation parameters are set as \(N=16\), \( K=12\), and \(L=2\). OFDM performance using the same number of null subcarriers and with distinct spacing \cite{ma2001non}, and the performance of the CP-based CFO estimation method \cite{zhang2021channel} are also plotted in Fig.~2. Evidently from Fig.~2, all three equalizers for OCDM-NSC collect full diversity, verifying Proposition 2 With this setup, OFDM asymptotically have unit diversity, whereas the performance of the CP-based CFO estimation method suffers because of the identifiability ambiguity. In this regard, we have testified that the proposed CFO estimator guarantees the CFO identifiability. Meanwhile, the null subchirps enable the multipath diversity of the OCDM-NSC system, and MLE and LEs collect the diversity gain. 

\begin{figure}
    \centering
    \includegraphics[width=8cm]{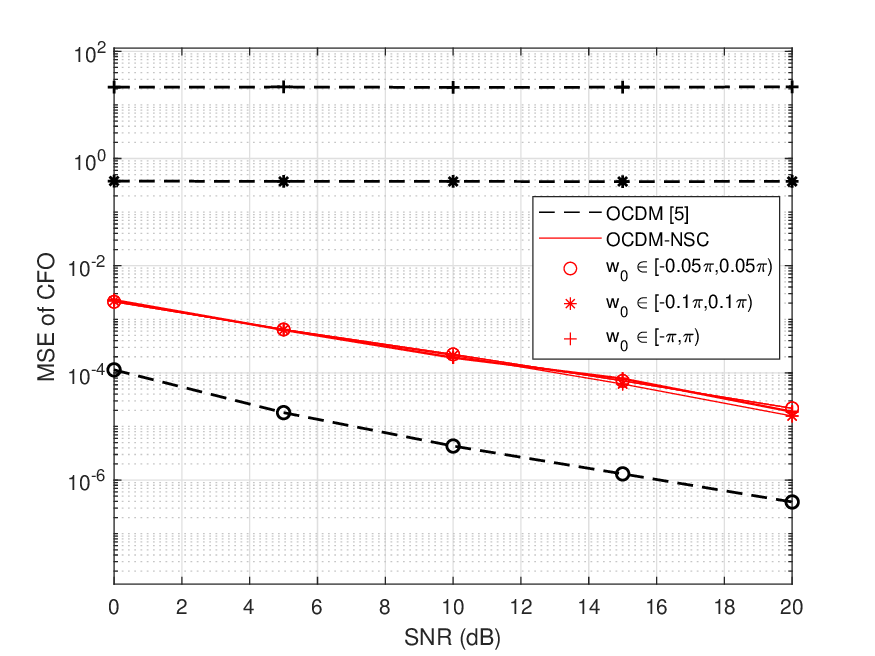}
    \caption{MSEs of CFO estimators versus SNR}
    \label{fig:2}
\end{figure}

\begin{figure}
    \centering
    \includegraphics[width=8cm]{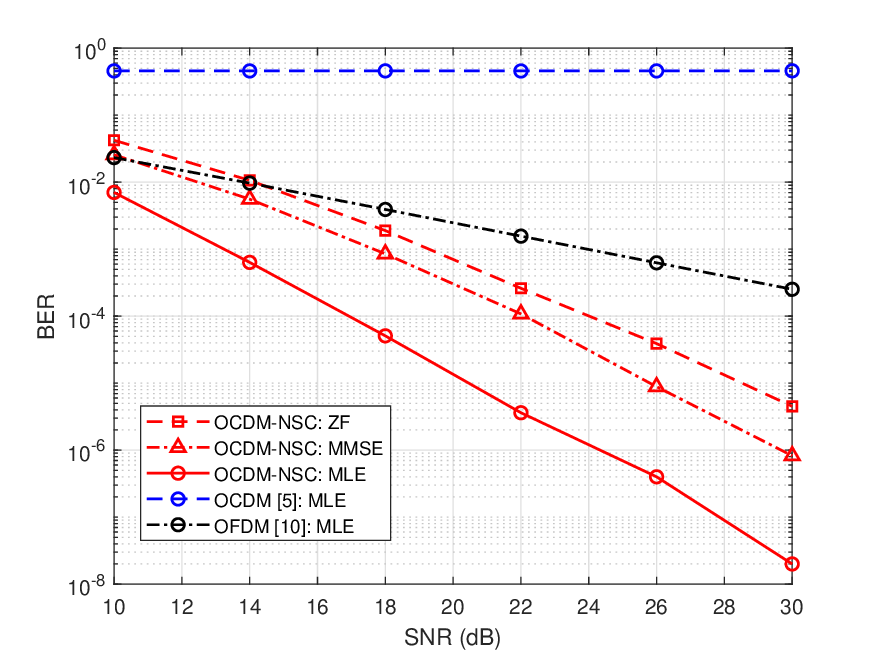}
    \caption{BER performance with the proposed CFO estimator}
    \label{fig:4}
\end{figure}

\section{Conclusion}
This paper investigates and solves the CFO identifiability problem in OCDM systems. 
We propose to insert consecutive null subchirps to facilitate the CFO estimation.  These null subchirps 
restore CFO identifiability by creating channel independent null subspace to overcome adverse channel conditions. A CFO estimator is accordingly proposed to achieve a full acquisition range. It has also been demonstrated that the OCDM system with consecutive null subchirps enables better performance than its plain counterpart. Finally, CFO identifiability for the proposed estimator is validated through simulations. Performance comparisons are made against other methods to show the advantages of the proposed OCDM-NSC. 

\bibliographystyle{IEEEtran}
\bibliography{references}

\begin{thebibliography}{10}
\providecommand{\url}[1]{#1}
\csname url@samestyle\endcsname
\providecommand{\newblock}{\relax}
\providecommand{\bibinfo}[2]{#2}
\providecommand{\BIBentrySTDinterwordspacing}{\spaceskip=0pt\relax}
\providecommand{\BIBentryALTinterwordstretchfactor}{4}
\providecommand{\BIBentryALTinterwordspacing}{\spaceskip=\fontdimen2\font plus
\BIBentryALTinterwordstretchfactor\fontdimen3\font minus \fontdimen4\font\relax}
\providecommand{\BIBforeignlanguage}[2]{{%
\expandafter\ifx\csname l@#1\endcsname\relax
\typeout{** WARNING: IEEEtran.bst: No hyphenation pattern has been}%
\typeout{** loaded for the language `#1'. Using the pattern for}%
\typeout{** the default language instead.}%
\else
\language=\csname l@#1\endcsname
\fi
#2}}
\providecommand{\BIBdecl}{\relax}
\BIBdecl

\bibitem{ouyang2016orthogonal}
X.~Ouyang and J.~Zhao, ``Orthogonal chirp division multiplexing,'' \emph{IEEE Transactions on Communications}, vol.~64, no.~9, pp. 3946--3957, 2016.

\bibitem{ouyang2017chirp}
X.~Ouyang, O.~A. Dobre, Y.~L. Guan, and J.~Zhao, ``Chirp spread spectrum toward the nyquist signaling rate—orthogonality condition and applications,'' \emph{ieee signal processing letters}, vol.~24, no.~10, pp. 1488--1492, 2017.

\bibitem{omar2019designing}
M.~S. Omar and X.~Ma, ``Designing ocdm-based multi-user transmissions,'' in \emph{2019 IEEE Global Communications Conference (GLOBECOM)}.\hskip 1em plus 0.5em minus 0.4em\relax IEEE, 2019, pp. 1--6.

\bibitem{ouyang2018robust}
X.~Ouyang, C.~Antony, G.~Talli, and P.~D. Townsend, ``Robust channel estimation for coherent optical orthogonal chirp-division multiplexing with pulse compression and noise rejection,'' \emph{Journal of Lightwave Technology}, vol.~36, no.~23, pp. 5600--5610, 2018.

\bibitem{zhang2021channel}
R.~Zhang, Y.~Wang, and X.~Ma, ``Channel estimation for ocdm transmissions with carrier frequency offset,'' \emph{IEEE Wireless Communications Letters}, vol.~11, no.~3, pp. 483--487, 2021.

\bibitem{de2022joint}
M.~L. de~Filomeno, L.~G. de~Oliveira, {\^A}.~Camponogara, A.~Diewald, T.~Zwick, M.~L. de~Campos, and M.~V. Ribeiro, ``Joint channel estimation and schmidl \& cox synchronization for ocdm-based systems,'' \emph{IEEE Communications Letters}, vol.~26, no.~8, pp. 1878--1882, 2022.

\bibitem{de2022discrete}
L.~G. de~Oliveira, B.~Nuss, M.~B. Alabd, A.~Diewald, Y.~Li, L.~Gehre, X.~Long, T.~Antes, J.~Galinsky, and T.~Zwick, ``Discrete-fresnel domain channel estimation in ocdm-based radar systems,'' \emph{IEEE Transactions on Microwave Theory and Techniques}, 2022.

\bibitem{omar2021performance}
M.~S. Omar and X.~Ma, ``Performance analysis of ocdm for wireless communications,'' \emph{IEEE Transactions on Wireless Communications}, vol.~20, no.~7, pp. 4032--4043, 2021.

\bibitem{trivedi2020low}
V.~K. Trivedi and P.~Kumar, ``Low complexity interference compensation for dfrft-based ofdm system with cfo,'' \emph{IET Communications}, vol.~14, no.~14, pp. 2270--2281, 2020.

\bibitem{ma2001non}
X.~Ma, C.~Tepedelenlioglu, G.~B. Giannakis, and S.~Barbarossa, ``Non-data-aided carrier offset estimators for ofdm with null subcarriers: identifiability, algorithms, and performance,'' \emph{IEEE Journal on selected areas in communications}, vol.~19, no.~12, pp. 2504--2515, 2001.

\bibitem{ghogho2001optimized}
M.~Ghogho, A.~Swami, and G.~B. Giannakis, ``Optimized null-subcarrier selection for cfo estimation in ofdm over frequency-selective fading channels,'' in \emph{GLOBECOM'01. IEEE Global Telecommunications Conference (Cat. No. 01CH37270)}, vol.~1.\hskip 1em plus 0.5em minus 0.4em\relax IEEE, 2001, pp. 202--206.

\bibitem{gao2006identifiability}
F.~Gao and A.~Nallanathan, ``Identifiability of data-aided carrier-frequency offset estimation over frequency selective channels,'' \emph{IEEE transactions on signal processing}, vol.~54, no.~9, pp. 3653--3657, 2006.

\bibitem{gao2008scattered}
F.~Gao, T.~Cui, and A.~Nallanathan, ``Scattered pilots and virtual carriers based frequency offset tracking for ofdm systems: algorithms, identifiability, and performance analysis,'' \emph{IEEE Transactions on Communications}, vol.~56, no.~4, pp. 619--629, 2008.

\bibitem{Wang_Optimality_2002}
Z.~Wang, X.~Ma, and G.~Giannakis, ``Optimality of single-carrier zero-padded block transmissions,'' in \emph{2002 IEEE Wireless Communications and Networking Conference Record. WCNC 2002 (Cat. No.02TH8609)}, vol.~2, 2002, pp. 660--664 vol.2.

\end{thebibliography}

\end{document}